\documentclass{article}

\usepackage[preprint, nonatbib]{neurips_2023}

\usepackage[utf8]{inputenc} 
\usepackage[T1]{fontenc}    
\usepackage{hyperref}       
\usepackage{url}            
\usepackage{booktabs}       
\usepackage{amsfonts}       
\usepackage{nicefrac}       
\usepackage{microtype}      
\usepackage{xcolor}         

\title{What Is AI Safety? What Do We Want It to Be?}

\author{%
  Jacqueline Harding\thanks{Equal contribution.}\\
  Stanford University\\
  \And
  Cameron Domenico Kirk-Giannini$^{*}$ \\
  Rutgers University--Newark\\
}

\begin{document}

\maketitle

\begin{abstract}
The field of AI safety seeks to prevent or reduce the harms caused by AI
systems. A simple and appealing account of what is distinctive of AI
safety as a field holds that this feature is constitutive: a research
project falls within the purview of AI safety just in case it aims to
prevent or reduce the harms caused by AI systems. Call this appealingly
simple account \emph{The Safety Conception of AI safety}. Despite its
simplicity and appeal, we argue that The Safety Conception is in tension
with at least two trends in the ways AI safety researchers and
organizations think and talk about AI safety: first, a tendency to
characterize the goal of AI safety research in terms of
\emph{catastrophic} risks from \emph{future} systems; second, the
increasingly popular idea that AI safety can be thought of as a branch
of \emph{safety engineering}. Adopting the methodology of conceptual
engineering, we argue that these trends are unfortunate: when we
consider what concept of AI safety it would be best to have, there are
compelling reasons to think that The Safety Conception is the answer.
Descriptively, The Safety Conception allows us to see how work on topics
that have historically been treated as central to the field of AI safety
is continuous with work on topics that have historically been treated as
more marginal, like bias, misinformation, and privacy. Normatively,
taking The Safety Conception seriously means approaching all efforts to
prevent or mitigate harms from AI systems based on their merits rather
than drawing arbitrary distinctions between them.
\end{abstract}

\section{Introduction}

As the development and deployment of artificial intelligence (AI)
proceed at breakneck speed, increasing attention is being paid to
\emph{AI safety}, an academic discipline which incorporates speculative
theoretical work on the alignment of advanced artificial systems with
human interests,\footnote{See, for example, Bostrom (2014), Russell
  (2019), and Ngo et al. (2023).} technical research on topics like
adversarial robustness,\footnote{Adversarial robustness studies, and
  aims at improving, the resilience of AI models against perturbations
  intended to yield unwanted outputs. See, for example, Carlini et al.
  (2019), Hendrycks and Dietterich (2019), and Bai et al. (2021).}
anomaly detection,\footnote{In this context, anomaly detection typically
  refers to methods for identifying inputs that fall outside of an AI
  system's intended range of application. See, for example, Chandola et
  al. (2009), Pang et al. (2021), and Yang et al. (2021).} model
interpretation,\footnote{Model interpretation refers to methods aimed at
  giving human-understandable insight into how AI models (especially
  deep learning models) work. See, for example, Elhage et al. (2021),
  Geiger et al. (2023), and, for a philosophical perspective on recent
  model interpretability work, Harding (2023).} the evaluation of
dangerous capabilities,\footnote{See, for example, Park et al. (2024),
  Shevlane et al. (2023), and Kinniment et al. (2023).} and
(increasingly) proposals concerning the governance of AI.\footnote{See,
  for example, Dafoe (2018), Brundage et al. (2020), Anderljung et al.
  (2023).} Despite its prominence in public discussions of AI, its
leaders' growing influence with policymakers, and its rise as a funding
area for grantmaking bodies, however, no consensus has emerged in
practice about what does or should mark a research topic as falling
within the purview of AI safety. In this paper, we aim to fill this
lacuna.

It's important to ward off a potential confusion at this stage. Although
we're arguing about what sort of research ought to be labeled as `AI
safety', we're not interested in terminology for terminology's sake. As
is well recognised by sociologists, how we draw the boundaries of an
academic discipline matters in many ways.\footnote{Paul Trowler, for
  example, notes that ``Disciplines are enacted as social practices are
  performed and as micropolitics are played out: teaching; research;
  conference attendance; departmental meetings; collaborative writing; mixed-disciplinary meetings of a political
  nature --- for example where resource allocation is at stake; funding
  applications, etc.'' (2012, 34). See also the other essays in Trowler,
  Saunders, and Bamber (2012).} For example, disciplinary boundaries
affect every aspect of research: they shape what researchers are
expected to read and engage with, who collaborates with whom, how norms
for the evaluation of research are set and which research trends are
taken seriously. Moreover, they affect the interaction between a
discipline and society more broadly: which researchers are viewed as
experts in the discipline in question, are included in public and
political conversations about the societal implications of research in
the discipline, and so on. Both categories of effects will be more
pronounced when disciplinary boundaries are more reified (in journals
and conferences, funding bodies, academic departments, etc.).\footnote{In
  what follows, we use the terms `discipline' and `field'
  interchangeably to refer to any recognizably cohesive community of
  researchers engaged in inquiry into a subject matter. We take it that
  it is clear that AI safety is a discipline in this sense. Some readers
  may wish to reserve the term `discipline' for research communities
  that have achieved a high level of institutional recognition, e.g. by
  being housed in their own academic departments. Such readers are
  welcome to replace our talk of disciplines with talk of fields.} So
the boundaries of the discipline of AI safety matter: they bear on how
research priorities within AI safety will be set, and on what sorts of
concerns and perspectives will be included in institutional discussions
of AI safety.

In what follows, our focus will be on one especially simple idea about
what AI safety is. According to this simple idea, which we call
\emph{The} \emph{Safety Conception of AI Safety} (or \emph{The Safety
Conception} for short), AI safety is the field of inquiry which seeks to
prevent or mitigate harms from AI systems. More precisely:

\begin{quote}
\textbf{The Safety Conception of AI Safety:} A research project belongs
to the field of AI safety just in case it is aimed at preventing or
reducing harms from AI systems' development and deployment.
\end{quote}

At first glance, The Safety Conception may seem too obvious or general
to be controversial. Indeed, many AI safety researchers would probably
express agreement with The Safety Conception if presented with it. But
we provide evidence below that members of the AI safety community often
speak and act in ways that are in tension with The Safety Conception. In
other words, while The Safety Conception may capture something close to
the \emph{manifest} concept of AI safety among AI safety researchers, it
does not capture the concept of AI safety that is \emph{operative} among
AI safety researchers or AI safety governance organizations.\footnote{The
  \emph{operative} concept of \emph{X} in a population is the concept
  which determines how members of the population apply \emph{X} in
  particular cases, while the \emph{manifest} concept of \emph{X} in a
  population is the concept which members of the population \emph{take}
  to determine how they apply X in particular cases. For further
  discussion, see Haslanger (1995, 2006).} For example, research on a
range of topics of interest to scholars in the Fairness, Accountability,
Transparency, and Ethics (\emph{FAccT} or \emph{FATE}) community,
including algorithmic bias,\footnote{See, for example, Buolamwini and
  Gebru (2018), Noble (2018), and Bellamy et al. (2019).}
misinformation,\footnote{See, for example, Bradshaw and Howard (2018),
  Keller and Klinger (2019), and Aïmeur et al. (2023).} breaches of
privacy,\footnote{See, for example, Allen (2016), Feldstein (2019), and
  Manheim and Kaplan (2019).} data theft,\footnote{This includes cases
  of artists' work being used as training data without their consent
  (see, for example, Chen (2023)).} distortions of the democratic
process,\footnote{See, for example, Howard et al. (2018), Manheim and
  Kaplan (2019), Helbing et al. (2019).} and private concentration of
power and resources,\footnote{See, for example, Nemitz (2018), Crawford
  (2021), and Bommasani et al. (2021).} clearly belongs to the field of
AI safety according to The Safety Conception. But research on these
topics is often treated as though it is not AI safety research or is
only marginally relevant to AI safety.

This situation is unfortunate, since we believe that taking The Safety
Conception more seriously would have significant benefits. Indeed, we
argue below that when we ask what operative concept of AI safety it
would be \emph{best} to have, there are compelling reasons to think that
The Safety Conception is the answer. In particular, The Safety
Conception has both explanatory and normative advantages. Explanatorily,
existing work in AI safety can be usefully viewed through the lens of
harm prevention or mitigation, and The Safety Conception allows us to
see how work on topics that have historically been treated as central to
the field of AI safety is continuous with work on topics that have
historically been treated as more marginal. Normatively, taking The
Safety Conception seriously means approaching all efforts to prevent or
mitigate harms from AI systems based on their merits rather than drawing
arbitrary distinctions between them. We argue that applying this
approach to research prioritization and AI governance is likely to
result in more effective harm reduction than alternatives.

Our thesis that The Safety Conception is the best conception of AI
safety has two main consequences:

\begin{quote}
(1): According to the best conception of AI safety, AI safety research
includes work on social harms from AI (such as bias and representational
harms, privacy and surveillance, and economic harms) as well as work on
`catastrophic' harms (such as enabling large-scale terrorism, state
warfare, or harms from autonomous, agentic AI).

(2): Given (1), there should be greater disciplinary integration between
researchers working on social harms and those working on catastrophic
harms. In particular, political conversations about AI safety should
include voices from researchers in both categories.
\end{quote}

Our argument in what follows is structured around our two primary goals:
first, demonstrating that The Safety Conception is not always the
conception of AI safety operative among AI safety researchers; second,
using the methodology of conceptual engineering to argue that it ought
to be. In Section 2, we introduce AI safety as a research area in more
detail and provide evidence that The Safety Conception is accepted by a
range of researchers at least as their manifest conception of AI safety.
In Section 3, we describe some ways in which the speech and actions of
AI safety researchers often come into conflict with The Safety
Conception in practice. In particular, we focus on the common ideas that
AI safety is especially or exclusively concerned with
\emph{catastrophic} risks from \emph{future} AI systems, and that it
should be construed as a branch of \emph{safety engineering}. In Section
4, we turn to the second of our two goals, introducing our conceptual
engineering methodology and identifying two purposes which a
characterization of AI safety should serve. In Section 5, we apply our
methodology to show that The Safety Conception best serves our two
purposes for a concept of AI safety. Section 6 concludes.

\section{The Safety Conception}

Something like The Safety Conception is ubiquitous in discussions by AI
safety researchers. Amodei et al. (2016), in an influential work which
served to set AI safety's early technical agenda, distinguish the field
of AI safety by its focus on reducing ``unintended and harmful
behavior that may emerge from poor design of real-world AI systems'' (p.
1). The Center for Security and Emerging Technology (CSET) defines AI
safety as ``an area of machine learning research that aims to identify
causes of unintended behavior in machine learning systems and develop
tools to ensure these systems work safely and reliably''.\footnote{\textless https://cset.georgetown.edu/publication/key-concepts-in-ai-safety-an-overview/\textgreater.}
The recently formed US AI safety Institute says that it ``exists to
help advance the understanding and mitigation of risks of advanced
AI''.\footnote{\textless https://www.nist.gov/system/files/documents/2024/05/21/AISI-vision-21May2024.pdf\textgreater.}
Even the Wikipedia entry on AI safety appears to endorse The Safety
Conception when it defines AI safety as ``an interdisciplinary field
focused on preventing accidents, misuse, or other harmful consequences
arising from AI systems''.\footnote{\textless https://en.wikipedia.org/wiki/AI\_safety\textgreater.}

If we endorse The Safety Conception, what does the research
landscape of AI safety look like? We can group research aimed at
preventing or mitigating harms from AI systems into three broad
categories: (i) non-empirical research on harms from AI systems, (ii)
empirical research on understanding current AI systems, (iii) empirical
research on preventing/mitigating harm from current AI systems.

\subsection{Non-empirical research on harms from AI systems}

There is a large body of research on the harms of AI systems that is
non-empirical. Much of this work can be thought of as aiming to identify
potential harms from AI systems that can then be assessed and tackled in
real systems.

Falling into this category are various taxonomies of harms from AI
systems, especially systems using generative models as
components.\footnote{See, for example, Weidinger et al. (2021),
  Weidinger et al. (2022), Shelby et al. (2023), and Solaiman et al.
  (2023).} For example, Weidinger et al. (2023) identify several broad
categories of harm: representational harms (including internal biases in
models as well as social-group-level disparities in AI system behavior),
toxicity harms (including the production of hateful content),
misinformation harms (when AI systems generate misleading information or
facilitate its dissemination), privacy harms (when private or sensitive
information in AI systems' training data can be extracted by users),
autonomy harms (when users become overly reliant on models), economic
harms (when automation exacerbates existing economic inequalities), and
environmental harms (such as the large energy cost of training and
deploying generative models).

As models become more powerful, researchers have argued that they will
create novel sorts of risk. Hendrycks et al. (2023) identify various
potential `catastrophic' risks (risks of large-scale harm) from advanced
AI systems, including enabling bioterrorism, cyberterrorism, state
misinformation and surveillance, and autonomous warfare. An influential
body of work also makes the case that advanced AI systems will pose
risks in their own right, rather than merely enabling or exacerbating
harms from other humans (e.g. Bostrom (2014), Carlsmith (2021),
Goldstein and Kirk-Giannini (2023), Bales et al. (2024)). Arguments for
this conclusion proceed in various ways; typically, they involve the
claim that as AI systems become more capable they will develop greater
degrees of agency, construed roughly as the ability to take sequences of
actions towards longer-horizon goals without human supervision. As AI
systems become more agentic, the argument goes, it will become likely
that their goals are in tension with human flourishing (since, e.g.,
they will `instrumentally converge' to goals involving the accumulation
of resources (Bostrom (2014))); this premise is concretized by reference
to examples of reward misspecification and goal misgeneralization in
reinforcement learning (e.g. Clark and Amodei (2016), Langosco et al.
(2022)).

Finally, note that much philosophical work on normative issues in AI's
development and deployment can also be understood as aimed at harm
reduction. This includes not only work on topics like algorithmic
discrimination and fairness, data theft, and privacy but also --- given
the arguments above --- work evaluating attributions of agency to AI
systems. It also includes research modeling different trajectories for
AI's development, involving predicting ways future systems could
develop, as well as proposals for the governance of future AI systems.

\subsection{Research on understanding current AI systems}

A prerequisite to preventing harms from AI systems' deployment is
understanding how the systems behave in a variety of deployment
settings. The nascent science of \emph{model evaluation} aims to measure
AI systems' capabilities (crudely, their performance ceiling on more
general cognitive tasks; see Harding and Sharadin (fc)) as well as the
ways these capabilities are mediated by different factors during
deployment, such as changes in the distribution of inputs. Model
evaluation often uses model-agnostic benchmarks, allowing standardized
comparisons between models.

Safety evaluation involves assessing the degree to which a system is
capable of producing harmful outputs; for an LLM, for example, this
might involve testing the model's ability to produce hate speech or
instructions to synthesize dangerous chemical compounds. Some safety
evaluation involves constructing novel benchmarks for dangerous
capabilities, whereas some involves more bespoke evaluation (Shevlane
et al. (2023)). Safety evaluation might also involve assessing for
harms which have occurred during the training of particular models, such
as environmental harms or harms to workers employed in the
data-processing pipeline. As systems improve, researchers have argued
that safety evaluation will involve testing for different sorts of
cognitive capabilities, such as those associated with systems' degree of
agency (including, for example, the degree to which models are
`self-aware' or pursue goals across a range of scenarios). There are
some early examples of benchmarks for these capabilities.\footnote{See,
  for example, Kinniment et al. (2023), Wijk et al. (2024).}

From a harm reduction perspective, one important aspect of model
evaluation is understanding the degree to which models' behavior can be
predicted in different deployment settings. This is often put in terms
of models' \emph{robustness} to changes in (e.g.) the sorts of inputs
they process, especially in adversarial settings in which an `attacker'
can perturb some aspect of the evaluation. Evaluating a system's degree
of adversarial robustness involves specifying a `threat model' for an
attacker (Carlini et al. 2019), which spells out her goals,
capabilities, and knowledge. \emph{Red-teaming} is a specific sort
of adversarial robustness evaluation in which the attackers' threat
model is chosen to closely mirror the environment in which the model
will be deployed (that is, the attacker plays the role of an
ill-intentioned user).\footnote{See, for example, Ganguli et al.
  (2022).}

Safety evaluation usually involves attempting to draw conclusions
about models from observing their behavior in different settings.
However, many researchers have argued that behavioral evidence will be
insufficient to provide the kind of safety guarantees needed, especially
as models exhibit higher degrees of agency (for example, many have
worried about deception, or notions of algorithmic discrimination which
are sensitive to \emph{how} the algorithm works, and not merely to its
input-output behavior).\footnote{See, for example, Park et al. (2024).}
Research on \emph{model interpretability} aims to solve this problem
by providing an understanding of how AI systems process inputs to
produce outputs, via observation of models' intermediate computations.
Older work on explainable AI tended to deliver relatively coarse-grained
understanding of models, such as visualization of which parts of the
input had the largest effect on the model's output. By contrast,
\emph{mechanistic interpretability}, an increasingly influential
subfield of machine learning, aims at giving a more complete causal
story about models' internal workings. Harding (2023) argues that
mechanistic interpretability delivers representational explanations of
model behavior, where the computation the model performs is expressed in
terms of manipulation of representations of abstract features of the
input. From a harm reduction perspective, the idea is that knowing the
`high-level algorithm' the model performs makes predicting (and
steering) its behavior easier.

\subsection{Research on preventing/mitigating harm from current AI systems}

Most of this research targets the model development process. This
includes research which investigates models' training data to uncover
training examples which perpetuate hateful content or contain dangerous
dual-use information. Training corpora can then be filtered to remove
these examples, and models which have been trained using these examples
can be flagged for potential safety concerns. It also includes research
on novel architectures which satisfy various safety-relevant properties,
such as transparency, as well as research on fine-tuning models on a
next-token-prediction task using supervised examples (SFT) or
reinforcement learning based on human or synthetic preference data over
model outputs (RLHF/AIF, or related non-RL tuning techniques, such as
direct preference optimization (DPO)).\footnote{See Christiano et al.
  (2017), Rafailov (2024).}

Assuming models' capabilities continue to improve, it will be
impractical (even, depending on the task, impossible) for humans to
supervise them directly. Indeed, the usefulness of RLHF derives from the
fact that --- for complex natural language tasks --- it is already
easier for humans to judge whether an output is desired than it is for
them to produce an example of a desired output. The research field of
\emph{scalable oversight} aims to develop techniques for supervising
models whose capabilities or knowledge exceeds those of the supervisor;
these are cases in which the supervisor can offer only ``weak
supervision''.\footnote{See, for example, Bowman et al. (2022).}
Various techniques have been proposed; one of the best known is
debate, in which --- for a given input (such as a question) --- the
supervisor observes an interaction (a \emph{debate}) between the AI
system and some other system (in practice, the other system is often a
copy of the original system), in which the AI system repeatedly attempts
to defend its output (such as the answer to the input question) against
challenges from the other system.\footnote{Irving, Christiano, and
  Amodei (2018).} The idea is that even if the supervisor is unable
to assess the quality of the AI system's initial output, she will be
able to judge the `winner' of the debate (since she will have observed
whether the output was able to be defended successfully); she will be
able to use this judgment to supervise the system.

Just as supervising models will become more difficult as models
develop, so too will assessing when supervision has been successful in
removing unwanted behaviors. This makes it challenging to evaluate
different proposals for scalable oversight. Bowman et al. (2022) propose
testing scalable oversight proposals using Cotra's (2021) `sandwiching'
paradigm. The idea is to test supervision techniques in domains in which
a model's capabilities lie in between (are `sandwiched' between) those
of a supervisor and an expert; the expert assesses the degree to which
the supervisor has been able to produce the desired behavior from the
model. The hope is that supervision techniques which are successful in
these domains will also be successful in domains in which no expert
exists.

There is also research aimed at mitigating harms during model
deployment. One sort of idea involves applying \emph{anomaly detection}
techniques to model inputs and outputs during deployment, to screen out
undesirable outputs as well as inputs likely to elicit them, such as
adversarial attacks (in LLMs, this manifests as content filters on
inputs and outputs).\footnote{See Pang et al. (2021) for an overview.}
Another involves inference-time interventions on models to steer
their outputs (in LLMs, this manifests as system prompts, or
interventions on the model's intermediate activations to make model
responses less harmful, guided by interpretability work).\footnote{See,
  for example, Li et al. (2024).} A third example involves
``watermarking'' model outputs, to mitigate downstream harms such as
plagiarism and disinformation.

Finally, an important category of research aimed at harm reduction
is explicitly \emph{sociotechnical}, in that it focuses on the
integration of AI systems within larger social systems. This includes,
for example, proposals for auditing AI systems throughout their
development and deployment.\footnote{See, for example, Raji et al.
  (2020), Costanza-Chock et al. (2022).} It also includes many proposals
for the governance of current AI systems.\footnote{See, for example,
  Dafoe (2018), Brundage et al. (2020), Anderljung et al. (2023).}
  
\subsection{Discussion}

There are two points to make about The Safety Conception at this
stage.

First, note that The Safety Conception does not distinguish between
different types of harms from AI systems when assessing whether research
counts as AI safety research; in particular, social harms (such as
representational harms) are treated the same way as `catastrophic'
harms, such as large-scale cybersecurity breaches. Work aimed at
reducing `existential' risks from AI systems' development and deployment
counts as AI safety research, but so too does much research involving
issues related to fairness, transparency, and accountability.

Second, one might worry that The Safety Conception is so liberal as
to be trivial. It is important to observe, then, that plenty of research
on AI-related normative issues does not count as AI safety research
under The Safety Conception. This is because there are many normative
dimensions to AI's development and deployment that have little to do
with harm reduction.

For example, much normative work in AI concerns questions about
\emph{power} and \emph{legitimacy} (Lazar (2022)); these questions are
not naturally posed in the language of harm reduction. Similarly, to the
extent we regard \emph{transparency} as an intrinsic value of systems
which make normatively significant decisions, work aimed at explaining
AI systems' behavior may be valuable even if it does not aim at reducing
harm.\footnote{For more on the value of transparency, see Grant et al.
  (2023).} When the field of algorithmic recourse uses counterfactual
explanation as a step in the process of delivering understanding (and,
possibly, a means of challenge) to users who have been negatively
affected by algorithmic decisions, then, it does not necessarily aim at
giving full explanations of model behavior in the sense that could be
used to improve model behavior (i.e. it does not aim at delivering
interventions to reduce future harm, unlike most work in mechanistic
interpretability).\footnote{See Verma et al. (2022).}

Although, as noted above, proposals for accountability might play a role
in AI governance proposals aimed at harm reduction, they are primarily
aimed at a different normative goal, that of \emph{justice} for those
harmed by AI systems. Indeed, current debates over whether `fairness' in
AI can be measured (e.g. Green and Hu (2018)) bear upon the question of
whether work on algorithmic fairness could ever be seen through the lens
of harm reduction.\footnote{This discussion illustrates a general
  pattern. Let X name some class of outcomes resulting from the
  development or deployment of an AI system. As the discussion above
  shows, on The Safety Conception, work in AI safety is largely aimed at
  answering the question `how can we prevent X from happening in the
  development and deployment of AI systems?'. Work on AI safety thus
  takes it for granted (a) that X is harmful (b) that, given enough
  information, we can measure whether an outcome in X occurs in a
  deployment scenario. Normative work that challenges these assumptions
  is often outside of the remit of AI safety.}

Furthermore, much normative work on AI relates to building systems
which promote human flourishing (broadly construed). For example,
researchers might be interested in how technological progress can enable
novel forms of political representation. This more `optimistic'
normative work falls outside of the remit of AI safety.

Finally, as the discussion above illustrates, some work on normative
questions in AI serves multiple purposes; The Safety Conception allows
for the possibility that work can fall within the remit of multiple
disciplines.

\section{Tensions with The Safety Conception}

So far, we have introduced The Safety Conception and provided evidence
that it is taken seriously by at least some AI safety researchers, at
least when it comes to their manifest concept of AI safety as a field of
study. In this section, we argue that even if The Safety Conception is
sometimes explicitly endorsed by AI safety researchers, in practice AI
safety as a research area often functions as though The Safety
Conception is false. In particular, we discuss (i) the tendency of some
AI safety researchers and organizations to explicitly or implicitly
restrict the domain of AI safety research so that it exclusively or
primarily concerns \emph{catastrophic} risk from \emph{future} systems,
and (ii) the common idea that AI safety should be understood as a branch
of \emph{safety engineering}, which imposes methodological constraints
on the kinds of research aimed at mitigating harms from AI systems that
can legitimately claim to belong to the field of AI safety.

\subsection{Catastrophic/Existential Harms}

The field of AI safety is often explicitly or implicitly defined so that
it exclusively or primarily concerns catastrophic harms from future AI
systems.\footnote{For the notion of a catastrophic or existential risk,
  see Ord (2020) and Bostrom and Ćirković (2008). For non-academic
  discussions of this way of thinking about AI safety, see Harding and
  Kirk-Giannini (2023), Roose (2023), and Schneier and Sanders (2023).}
We will refer to this way of defining AI safety as \emph{The
Catastrophic Conception}. The Catastrophic Conception has been around as
long as the phrase ``AI safety'' itself,\footnote{The Machine
  Intelligence Research Institute (MIRI), an organization that was
  influential in AI safety's development as a research field, explains
  the need for AI safety research as follows: ``If {[}an AI{]} system's
  assigned problems/tasks/objectives don't fully capture our real
  objectives, it will likely end up with incentives that
  catastrophically conflict with what we actually want\ldots{} our
  take-away from this is that we should prioritize early research into
  aligning future AI systems with our interests.''} and it persists
today. Indeed, a recent study of the `epistemic community' of AI safety
characterizes the discipline as follows: ``generally, AI safety
practitioners are interested in preventing \emph{catastrophic long-term
events} precipitated by the deployment of machine learning systems''
(Ahmed et al. 2024, emphasis added). Dalrymple et al. (2024), whose
authors include several prominent AI researchers, describe ``The AI
Safety Problem'' as the claim that ``sufficiently advanced AI systems
may threaten the survival of the human species, or lead to our permanent
disempowerment, especially in the case of AI systems that are more
intelligent than humans.'' (p. 2).

Definitions of AI safety that prioritize catastrophic/existential risks
appear even in sources that may initially appear to endorse The Safety
Conception. For example, the Wikipedia article on AI safety, cited above
for describing AI safety as ``an interdisciplinary field focused on
preventing accidents, misuse, or other harmful consequences arising from
AI systems,'' goes on to say that ``the field is particularly concerned
with existential risks posed by advanced AI models,'' with an in-text
hyperlink to the article ``Existential risk from AI.''\footnote{\textless https://en.wikipedia.org/wiki/AI\_safety\textgreater.}
Similarly, in an influential recent paper on AI safety,
researchers from UC Berkeley, Google, and OpenAI write, ``We define ML
Safety research as ML research aimed at making the adoption of ML more
beneficial, \emph{with emphasis on long-term and long-tail risks}''
(Hendrycks et al. 2021, emphasis added).\footnote{In this context, the
  expression ``long-tail risks'' should be understood as roughly
  synonymous with ``catastrophic risks.''}

Similar characterizations of AI safety are offered in the
self-descriptions or mission statements of various organizations active
in the area. The AI Alignment Forum, a website used by AI safety
researchers to discuss technical and theoretical developments in the
field, describes the reason for its creation as: ``Foremost, because
misaligned powerful AIs may pose the greatest risk to our civilization
that has ever arisen.''\footnote{\textless https://www.alignmentforum.org/posts/Yp2vYb4zHXEeoTkJc/welcome-and-faq\textgreater.}
The same website highlights a series of posts intended as an
introduction to the alignment problem written by Richard Ngo, an AI
safety researcher who has worked at OpenAI and DeepMind. Ngo begins:

\begin{quote}
``The key concern motivating technical AGI safety research is that
we might build autonomous artificially intelligent agents which are much
more intelligent than humans, and which pursue goals that conflict with
our own\ldots{} AIs will eventually become more capable than us at the
types of tasks by which we maintain and exert that control. If they
don't want to obey us, then humanity might become only
Earth's second most powerful ``species'', and lose the
ability to create a valuable and worthwhile future.'' (Ngo 2020)
\end{quote}

We read Ngo's use of the phrase `AGI safety' here as a way of signaling
that he embraces The Catastrophic Conception and rejects The Safety
Conception. That is, we take it that he is \emph{not} intending to
commit to the existence of two disciplines, AI safety and AGI safety
(with the latter perhaps a subdiscipline of the former). Instead, he is
using `AGI safety' to pick out the same discipline others refer to as
`AI safety' or `ML safety', and claiming that it is motivated by a
concern with future catastrophic risks. This way of using terminology is
also adopted by Bowman (2022), in a document providing a high-level
overview of the discipline of AI safety, who uses `AI safety' and `AGI
safety' interchangeably, whilst acknowledging that the phrase `AI
safety' is sufficiently vague to permit other interpretations (such as,
presumably, something like The Safety Conception). Note again that our
point here is not merely semantic; we will argue that the best
conception of AI safety does not draw a sharp disciplinary boundary
between catastrophic/existential and non-catastrophic risks. So our
arguments are intended to apply to any view which attempts to carve out
a separate discipline which deals only with existential risks from
advanced AI, regardless of what label it gives this discipline.

In implicitly identifying the field of AI safety with the more
specific concerns of those interested in AGI rather than AI in general,
and in motivating the need for AI safety research by appealing to the
possibility of long-term catastrophic outcomes like civilizational
collapse and human disempowerment, these researchers and organizations
de-center a range of research topics which clearly fall within the field
of AI safety on The Safety Conception, including bias, toxicity,
misinformation, privacy harms, economic harms, and environmental harms,
as well as technical research on how to reduce the short-term harms of
near-future AI systems.

\subsection{Safety Engineering}

A second and quite different way in which the practice of AI safety
researchers comes into conflict with The Safety Conception is that AI
safety is sometimes understood to be a largely technical branch of
\emph{safety engineering}. Safety engineering is an interdisciplinary
field which aims to reduce harm from the development and deployment of
engineered systems (Roland and Moriarty (1990)). Like many engineering
fields, safety engineering is best characterized by its objectives and
methodologies rather than by the questions it seeks to answer. Broadly
speaking, the practice of safety engineering takes as its object of
study engineered systems and the environments in which they are
developed and deployed, which themselves can be thought of as
sociotechnical systems. It takes as its goal the minimization of harm
from the development and deployment of these systems and involves
activities like hazard identification (identifying cases in which a
system's deployment leads to unwanted outcomes), risk analysis
(identifying how likely each identified hazard is to occur), and risk
management (providing suggestions for reducing the aggregate risk from a
system's deployment).

Several authors have suggested that ``AI safety'' should be viewed as
continuous with the larger field of safety engineering.\footnote{See,
  amongst others, Hutchins (1995), Yampolskiy and Fox (2013), Dobbe
  (2022), Weidinger et al. (2023), Rismani, Shelby, Smart, Delos Santos,
  et al. (2023), Rismani, Shelby, Smart, Jatho, et al. (2023), Fang and
  Johnson (2019), Hendrycks et al. (2022), Khlaaf et al. (2022),
  Koessler and Schuett (2023), Trapp et al. (2018), Raji et al. (2020),
  and Costanza-Chock et al. (2022).} Different authors make different
claims here; some authors merely claim that lessons from safety
engineering should be applied to the deployment and development of AI.
We are interested in a stronger version of the claim, which takes
continuity with safety engineering as constitutive of AI safety as a
discipline. This view is suggested, for example, by Weidinger et al.
(2023) when they advocate a sociotechnical approach to AI safety
''inspired by a system safety approach from the discipline of safety
engineering'' (p. 8) and then remark:

\begin{quote}
``current and future\ldots{} classes of generative AI systems have been
claimed to possess novel capabilities that may create `extreme' risks to
society, such as from disseminating dangerous information or creating
novel types of cyber attacks... Historically, these focus areas --- or
ethical and safety risks --- associated with AI systems have been
fragmented and have constituted distinct research communities based on
perceived epistemic differences and differences in timely proximity of
harms... However, recent advances in generative AI systems are forcing a
collapse of these epistemological silos\ldots{} The sociotechnical
approach put forward here accommodates risks that are of concern to both
research communities and it can thus serve to coordinate work between
these communities on risks from generative AI systems.''
\end{quote}

For Weidinger et al., the AI ethics and AI safety research communities
are united by their subsumption within the larger discipline of
(sociotechnical) safety engineering. Let us call this \emph{The
Engineering Conception} of AI safety.

What does The Engineering Conception of AI safety amount to? That is,
what does it mean to claim that AI safety is continuous with safety
engineering? It can't just be that research in AI safety, like research
in safety engineering more broadly, aims at harm prevention and
mitigation. If this were all The Engineering Conception consisted in, it
would collapse into The Safety Conception --- the connection to safety
engineering would be entirely epiphenomenal. To have content, the claim
must be that (a) there is some set of methodologies which is distinctive
of the discipline of safety engineering (b) work falls within AI safety
to the extent it can be understood as applying these methodologies to
the domain of AI.

And indeed, many technical fields in contemporary AI safety can be seen
as examples of safety engineering. For example, methods relating to
distribution-shift and adversarial robustness, as well as anomaly and
trojan detection and calibration, can be understood as efforts to
estimate and improve the reliability of different components of AI
systems.\footnote{For further discussion, see Corso et al. (2023).}
Similarly, work on reward misspecification and goal misgeneralization in
reinforcement learning (e.g. Clark and Amodei (2016), Langosco et al.
(2022)) can be understood as hazard identification and analysis.

Yet conceiving of AI safety as a branch of safety engineering is in
tension with The Safety Conception. There are many research projects
that aim to prevent or mitigate harms from AI systems which do not
recognizably employ the methodology of engineering. These include, for
example, theoretical work involving premises like the orthogonality
thesis and the idea of instrumental convergence (e.g. Bostrom (2014)),
attempts to quantify the risks posed by these kinds of issues (e.g.
Carlsmith (2021), Goldstein and Kirk-Giannini (2023)), conceptual work
on the alignment problem (Gabriel (2020)) and related issues in
normative theory (e.g. D'Alessandro (2024), Tubert and Tiehen (2024),
Thornley (2024)), and governance proposals made by AI safety
researchers, such as licensing regimes for models based on their
training compute budgets (Shavit (2023)). The Engineering Conception of
AI safety excludes these projects, whereas The Safety Conception does
not.

\section{Conceptual Engineering}

We turn now from showing that The Safety Conception is non-trivial in
the sense that taking it seriously would require reconsidering certain
trends in the way AI safety researchers and organizations think and talk
about AI safety to arguing that The Safety Conception is the conception
of AI safety we \emph{ought} to adopt, and in particular to arguing for
claims (1) and (2), reproduced below:

\begin{quote}
(1): According to the best conception of AI safety, AI safety research
includes work on social harms from AI (such as bias and representational
harms, privacy and surveillance, and economic harms) as well as work on
`catastrophic' harms (such as enabling large-scale terrorism, state
warfare, or harms from autonomous, agentic AI).

(2): Given (1), there should be greater disciplinary integration between
researchers working on social harms and those working on catastrophic
harms. In particular, political conversations about AI safety should
include voices from researchers in both categories.
\end{quote}

Our methodology in arguing for these claims is one of conceptual
engineering: the project, as Cappelen puts it, of ``assessing and
improving our representational devices'' (2018, p. 3). We introduce our
conceptual engineering project in more detail in this section before
defending our claim that The Safety Conception is the best conception of
AI safety in section 5 below.

In evaluating various candidate concepts of AI safety, we will be
interested both in the extent to which each is explanatorily useful and
the extent to which it helps to promote the practical goal of reducing
harms from AI systems. Assessing concepts along the first of these
dimensions is standard practice in conceptual engineering; it is, for
example, the approach taken by Clark and Chalmers (1998) in arguing for
the thesis that mental states and cognitive processes can extend beyond
the boundaries of the brain, which Cappelen (2018) cites as an early
paradigm of conceptual engineering.\footnote{Other recent applications
  of the methodology of conceptual engineering for explanatory utility
  include Tanswell (2018), Isaac (2020), and Kirk-Giannini (2023, fc).}

Assessing concepts according to the extent to which they promote
practical goals is also a form of conceptual engineering, but one which
has usually been discussed using the term \emph{ameliorative inquiry}
(Haslanger (2005)).\footnote{It is sometimes also called
  \emph{analytical inquiry} (Haslanger (2000)).} Ameliorative inquiry is
a common methodology in the feminist tradition, which seeks to construct
emancipatory accounts of social categories like gender and sexual
orientation. As Dembroff (2016, 4) clarifies, the methodology of
ameliorative inquiry has two important components: ``Elucidating
purposes ideally served by our {[}target{]} concept,'' and
``Re-engineering our {[}target{]} concept\ldots{} in light of
{[}these{]} purposes.'' Our discussion in what follows will be
structured around these two components.

What, then, are the purposes ideally served by our concept of AI safety?
We will focus on two:

\begin{quote}
(A): It should provide a unifying explanation of what makes it the case
that paradigmatic research programs in AI safety belong to the same
field of inquiry.

(B): It should be conducive to reducing the harms caused by AI systems.
\end{quote}

The first of these purposes speaks to the explanatory utility of a
concept of AI safety. A concept that strays too far from the
conventional understanding of what is included in the field risks simply
changing the subject. At the same time, we must be open to the
possibility that the best account of what is distinctive about AI safety
as a field of inquiry will issue some surprising verdicts: some research
questions which have not traditionally been regarded as AI safety
research questions might turn out to fall within the purview of AI
safety, and some research questions which have traditionally been
regarded as AI safety research questions might turn out not to be.

The second purpose speaks to the practical implications of a concept of
AI safety. In proposing an ameliorative concept of AI safety, we do not
claim, implausibly, that anyone is directly helped or harmed by any
concept, or that the harms caused by AI systems are directly mediated by
the representational devices we use to individuate fields of inquiry.
Instead, our claim is that the way in which we think about fields of
inquiry has indirect effects on the amount of harm caused by AI systems,
mediated by decisions about how to prioritize research projects across
areas of inquiry and choices about which experts' input is given
consideration in crafting policy. If there are areas of inquiry which
would, if adequately prioritized and given a voice in policy
discussions, reduce the harms caused by AI systems, and if these areas
of inquiry are deprioritized and ignored in AI safety policy discussions
because of the concept of AI safety currently operative among AI
researchers, we regard this as a practical reason to engage in
ameliorative revision of the concept.

\section{Conceptual Engineering Supports The Safety Conception}

Why think that The Safety Conception is the best way of demarcating AI
safety as a research area? In this section, we argue that The Safety
Conception accomplishes our purposes (A) and (B) better than alternative
proposals. With respect to each purpose, our argument will have the
following form: First, we will argue that The Safety Conception does a
good job of serving that purpose. Second, we will compare The Safety
Conception with The Catastrophic Conception and The Engineering
Conception, arguing that The Safety Conception fares better than these
alternatives. Third, we will offer some considerations which lead us to
generalize to the conclusion that \emph{any} departure from The Safety
Conception will yield an understanding of AI safety that fares worse
with respect to that purpose.

Consider first purpose (A): A satisfactory concept of AI safety should
provide a unifying explanation of what makes it the case that
paradigmatic research programs in AI safety belong to the same field of
inquiry. Providing a unifying explanation of this kind is not a trivial
achievement, since it is not immediately apparent why (for example)
philosophical work on whether agents in general might have instrumental
reasons to seek power should fall within the same field of inquiry as
technical machine learning work on anomaly detection. The Safety
Conception explains what ties these kinds of research questions together
into a single area of inquiry: they aim at preventing or mitigating
harms from AI systems. Indeed, in our view The Safety Conception
embodies the ideal level of generality for thinking about AI safety. It
provides a unifying explanation of what makes it the case that
paradigmatic research programs in AI safety belong to the same field of
inquiry while also highlighting the continuity between those research
programs and topics like algorithmic bias, misinformation, and
distortions of the democratic process, which have historically be
treated as marginal AI safety research topics if they have been
understood to fall within the domain of AI safety research at all.

So The Safety Conception does a good job of serving purpose (A). But our
claim is stronger --- that The Safety Conception does better than
alternatives at serving purpose (A). Consider, then, how The Safety
Conception compares in this context to The Catastrophic Conception. In
our view, the latter conception of AI safety fares poorly with respect
to purpose (A). To see this, note that, like work in the wider
contemporary machine learning landscape, most technical work in AI
safety is empirical, based on experiments on existing models. Central
questions in AI safety --- for example, questions about adversarial
robustness, model fine-tuning using human preference data, and
interpretability --- concern present systems just as much as future ones
and have no deep conceptual connection to issues specifically of
catastrophic or existential risk --- they appear largely agnostic to the
kinds of risks to which they are applied. So a concept of AI safety
which tied it constitutively to catastrophic future risks would
arbitrarily exclude a great deal of paradigmatic AI safety work focused
on mitigating near-term non-catastrophic harms.

This exclusion would be especially unfortunate given the significant
continuities between non-catastrophic and catastrophic risks from AI.
For example, language models produce hate speech for the same reason
they produce instructions on making bombs (properties of their
pre-training corpora), and we miss something important from the
perspective of explanation when we ignore this. Indeed, many of the
concrete catastrophic risks identified by AI safety researchers (e.g. by
Hendrycks et al. (2023)), such as AI enabling permanent political
disempowerment by undermining democratic processes, are simply `scaled
up' versions of risks already well-studied by researchers focusing on
social harms from AI.\footnote{On this subject, see also Kasirzadeh
  (2024).}

Similar remarks apply to The Engineering Conception. Many central
research projects in AI safety, such as the project of assessing whether
intelligent artificial agents are likely to have instrumental reasons to
act in ways that harm humans, are not best approached using the tools or
methods of engineering. Conceiving of AI safety as a kind of engineering
would arbitrarily exclude research on these topics.

These remarks about the Catastrophic Conception and the Engineering
Conception lead us to believe that no conception of AI safety more
restrictive than The Safety Conception could fare better than The Safety
Conception with respect to purpose (A) --- narrowing the purview of AI
safety risks losing out on explanatorily important continuities between
different research programs. At the same time, we also believe that any
\emph{less} restrictive conception would fare poorly with respect to
purpose (A) as compared to The Safety Conception. Such a conception
would, by stipulation, include in the domain of AI safety some research
projects not aimed at preventing or mitigating harms from AI systems.
And it is difficult to see how the resulting collection of research
projects could be explanatorily unified. It follows that The Safety
Conception describes the best concept of AI safety when it comes to
purpose (A).

Consider now purpose (B): A satisfactory concept of AI safety should be
conducive to reducing the harms caused by AI systems. Our argument that
The Safety Conception fares well with respect to purpose (B) is
structured around a comparison between three possible ways of
structuring the AI safety research community:

\begin{quote}
In the first, which we might call \emph{The Safe Scenario}, the AI
safety research community is structured in accordance with The Safety
Conception. Research projects are prioritized based on their expected
contribution to the goal of preventing or mitigating harms from AI
systems, and experts are consulted during political deliberation about
AI safety to the extent that they are knowledgeable about how to prevent
or mitigate harms from AI systems.

In the second, which we might call \emph{The Catastrophic Scenario}, the
AI safety research community is structured in accordance with The
Catastrophic Conception. Research projects are prioritized based on
their expected contribution to the goal of preventing or mitigating
catastrophic harms from future AI systems, and experts are consulted
during political deliberation about AI safety to the extent that they
are knowledgeable about how to prevent or mitigate catastrophic harms
from future AI systems.

In the third, which we might call \emph{The Engineering Scenario}, the
AI safety research community is structured in accordance with The
Engineering Conception. Research projects are prioritized based on their
expected contribution to the goal of preventing or mitigating harms from
AI systems, subject to the constraint that they employ methods
recognizable as engineering, and experts are consulted during political
deliberation about AI safety to the extent that they are engineers
knowledgeable about how to prevent or mitigate harms from AI systems.
\end{quote}

To begin, note that in The Safe Scenario, the AI safety community is
likely to be quite effective at reducing the harms caused by AI systems.
This is because it prioritizes research projects solely based on their
expected contribution to the goal of preventing or mitigating harms from
AI systems and consults experts solely to the extent that they are
knowledgeable about how to prevent or mitigate harms from AI systems.

Now contrast The Safe Scenario with The Catastrophic Scenario and The
Engineering Scenario. In both The Catastrophic Scenario and The
Engineering Scenario, the focus of the AI safety community is restricted
in some way: in the former case, to catastrophic harms from future AI
systems, in the latter case, to harm reduction efforts that employ
engineering methods. We think that both kinds of restrictions are likely
to make the AI safety community less effective at reducing the harms
caused by AI systems. In The Catastrophic Scenario, there are some
research projects focused on reducing the harms caused by AI systems ---
namely, those research projects which target present and/or
non-catastrophic harms --- which are automatically deprioritized.
Similarly, there are some experts --- namely, those experts who are
knowledgeable about how to prevent or mitigate present and/or
non-catastrophic harms --- who will be excluded from political
deliberation about AI safety. The same worry applies, mutatis mutandis,
in The Engineering Scenario: research projects which fall outside the
disciplinary boundaries of engineering will automatically be
deprioritized, and experts knowledgeable about such research projects
will be excluded from political deliberation about AI safety.

These differences from The Safe Scenario are likely to make the AI
safety community less effective at reducing the harms caused by AI
systems because they move it away from the practice of prioritizing
research directions solely according to their merit \emph{qua} harm
reduction effort and consulting experts solely according to their
expertise when it comes to harm reduction efforts. The focus on future
catastrophic risks in The Catastrophic Scenario ignores the significant
continuities between catastrophic and non-catastrophic risks from AI.
This oversight strikes us as net safety-negative, since it is plausible
that allocating AI safety resources to researchers with broader
sociotechnical expertise will lead to new insights and proposals
concerning the whole spectrum of risks, including catastrophic risks.

Another point to be made in this connection is that it is much easier to
do --- and, importantly, assess --- technical work on systems that
actually exist.\footnote{This is not to say that it is impossible to do
  empirical work aimed at future systems; see, e.g., our discussion of
  `scalable oversight' above.} Construing AI safety as concerned in the
first instance only with catastrophic harms from future systems means
that work on present systems can be justified as AI safety work only if
it can be shown to reduce catastrophic risks from future systems. Any
attempt to make this case will rely on an auxiliary premise, namely that
there will be sufficient continuities between present and future systems
that lessons learned from experiments on the former will apply to the
latter (e.g. that effective oversight strategies on models which produce
text-only outputs will continue to be effective on different classes of
models, or that current insights from mechanistic interpretability will
generalize beyond the transformer architecture).

Regardless of the plausibility of this premise (which will vary from
case to case), it is independent of the actual research contribution
made by experiments on present systems; two papers could perform similar
sets of experiments, but only one could frame their contribution in
terms of future systems. In practice, the researchers who will make this
auxiliary premise explicit (i.e. who directly attempt to connect their
contribution to reducing harms from future, more capable systems) will
be precisely those who have already bought into The Catastrophic
Conception. This kind of gatekeeping strikes us as likely to make the
discipline of AI safety less effective at reducing the harms caused by
AI systems.

Finally, there is a risk in The Catastrophic Scenario that the idea that
AI safety only relates to catastrophic harms from future systems will
enable model developers to talk about safety while resisting effective
regulatory proposals (i.e. `safety-washing' (Perrigo (2023)).

Similar worries arise about The Engineering Scenario. Automatic
deprioritization of research projects that do not adopt the methodology
of safety engineering is likely to lead to less safe outcomes. And model
developers could espouse a commitment to AI safety while ignoring
safety-critical theoretical or sociotechnical research projects that do
not fall within the disciplinary boundaries of engineering. These
include in particular harms from autonomous, agentic AI, for which no
analogue exists in traditional safety engineering.

There is a more general point to be made here, beyond the claim that the
AI safety community in The Safe Scenario is likely to be more effective
than the AI safety communities in The Catastrophic Scenario and The
Engineering Scenario: \emph{any} way of choosing research priorities or
selecting which experts to consult in political deliberation about AI
safety other than the one embodied in The Safe Scenario is likely to be
less effective at reducing harm. In so far as The Safe Scenario is the
scenario that embodies The Safety Conception, we have reason to believe
that The Safety Conception is the best conception of AI safety for
purpose (B).

We have argued that there are reasons for thinking that The Safety
Conception does better than competitors when it comes to purpose (A),
and also that it does better than competitors when it comes to purpose
(B). It follows that The Safety Conception is the best conception of AI
safety.

Before concluding, it is worth addressing a possible worry about our
argument that The Safety Conception fares better than competitors with
respect to reducing harms from AI systems. In particular, some might
worry that the Safe Scenario is likely to be less safe than the
Catastrophic Scenario because in the Safe Scenario, research directions
aimed at preventing catastrophic harms from AI systems will be less
prioritized.

It is important to realize that the fact that the AI safety community is
not exclusively concerned with catastrophic harms in The Safe Scenario
does not entail that work on preventing or mitigating catastrophic harms
will be less prioritized. The situation is rather that in The Safe
Scenario work focused on catastrophic harms will not be prioritized
\emph{automatically}. If efforts to prevent or mitigate catastrophic
harms from AI systems have more merit \emph{qua} harm reduction effort
than other kinds of interventions, then they will be prioritized in The
Safe Scenario as they are in The Catastrophic Scenario. Conversely, if
it turns out that efforts to prevent or mitigate catastrophic harms do
not have more merit \emph{qua} harm reduction effort than other kinds of
interventions (as we suspect it will), it seems to us that the proper
conclusion to draw is that The Catastrophic Scenario is likely to be
less safe than The Safe Scenario.

\section{Conclusion}

AI systems are potentially dangerous in myriad ways, and it is of
central importance in deploying them to think carefully about how to
prevent or mitigate the harms they can cause. This is the basic premise
of AI safety research.

We have argued that this basic premise also picks out the best
conception of AI safety as a field: The Safety Conception. When we think
carefully about how to demarcate the field of AI safety in a way that is
explanatorily fruitful and mitigates the harms AI systems can cause, it
becomes clear that The Safety Conception is superior to rival proposals
like The Catastrophic Conception and The Engineering Conception. It
follows that the Safety Conception ought to be the operative concept of
AI safety among AI safety researchers and policymakers, not merely the
manifest concept.

Concretely, this means that research on social harms from AI should be
presented at and published in the same venues as research on
catastrophic harms; if existing conferences and publication venues
cannot accommodate this, new ones which can should be created. It means
that researchers on LLM toxicity should be following developments in
mechanistic interpretability, and that researchers on AI deception
should draw on sociotechnical work on misinformation. It means that AI
labs should not have separate ``ethics'' and ``safety'' teams, and that
AI safety funders should be open to funding research which does not
explicitly frame its contribution in terms of reducing catastrophic or
existential risks.

These suggestions, we anticipate, will strike many readers as obviously
sensible, continuous with many proposals to broaden AI safety's tent in
recent years (Lazar and Nelson 2023). We take this to be a virtue of The
Safety Conception: it serves to motivate and justify common sense
recommendations for disciplinary integration between those working on a
large array of different harms from AI.

\section*{Acknowledgements}

Thanks especially to Seth Lazar for detailed comments on an earlier draft.

\section*{References}

{
\small

Ahmed, Shazeda, Klaudia Jaźwińska, Archana Ahlawat, Amy Winecoff, and
Mona Wang. `Field-Building and the Epistemic Culture of AI Safety'.
\emph{First Monday}, 14 April 2024.
\textless https://doi.org/10.5210/fm.v29i4.13626 \textgreater.

Aïmeur, E., Amri, S., and Brassard, G. (2023). Fake news, disinformation
and misinformation in social media: A review. \emph{Social Network
Analysis and Mining} 13(30): 1--36.

Allen, A. L. (2016). Protecting one\textquotesingle s own privacy in a
big data economy. \emph{Harvard Law Review Forum} 130: 71--78.

Amodei, D., Olah, C., Steinhardt, J., Christiano, P., Schulman, J., \&
Mané, D. (2016). Concrete problems in AI safety. ArXiv preprint.
\textless https://arxiv.org/abs/1606.06565\textgreater.

Anderljung, M., Barnhart, J., Korinek, A., Leung, J.,
O\textquotesingle Keefe, C., Whittlestone, J., Avin, S., Brundage, M.,
Bullock, J., Cass-Beggs, D., Chang, B., Collins, T., Fist, T., Hadfield,
G., Hayes, A., Ho, L., Hooker, S., Horvitz, E., Kolt, N. \ldots{} and
Wolf, K. (2023). Frontier AI regulation: Managing emerging risks to
public safety. ArXiv preprint.
\textless https://arxiv.org/abs/2307.03718\textgreater{}

Bai, T., Luo, J., Zhao, J., Wen, B., and Wang, Q. (2021). Recent
advances in adversarial training for adversarial robustness.
\emph{International Joint Conference on Artificial Intelligence
(IJCAI-21).}

Bales, A., D\textquotesingle Alessandro, W., \& Kirk‐Giannini, C. D.
(2024). Artificial intelligence: Arguments for catastrophic risk.
\emph{Philosophy Compass}, 19(2): e12964.

Bellamy, R.K., Dey, K., Hind, M., Hoffman, S.C., Houde, S., Kannan, K.,
Lohia, P., Martino, J., Mehta, S., Mojsilović, A., Nagar, S.,
Ramamurthy, K. N., Richards, J., Saha, D., Sattigeri, P., Singh, M.,
Varshney, K. R., and Zhang, Y. (2019). AI Fairness 360: An extensible
toolkit for detecting and mitigating algorithmic bias. \emph{IBM Journal
of Research and Development} 63(4/5), 4: 1--15.

Bommasani, R., Hudson, D.A., Adeli, E., Altman, R., Arora, S., von
Arx, S., Bernstein, M.S., Bohg, J., Bosselut, A., Brunskill, E.,
Brynjolfsson, E., Buch, S., Card, D., Castellon, R., Chatterji, N.,
Chen, A., Creel, K., Davis, J. Q., Demszky., D. \ldots{} and Liang, P.
(2021). On the Opportunities and Risks of Foundation Models. ArXiv
preprint \textless https://arxiv.org/abs/2108.07258 \textgreater

Bostrom, N. (2014). \emph{Superintelligence: Paths, Dangers,
Strategies}. Oxford University Press.

Bostrom, N. and Ćirković, M. M. (eds.) (2008). \emph{Global Catastrophic
Risks}. Oxford University Press.

Bowman, S. (2022). `AI Safety and Neighboring Communities: A Quick-Start
Guide, as of Summer 2022', 1 September 2022.
\textless https://www.lesswrong.com/posts/EFpQcBmfm2bFfM4zM/ai-safety-and-neighboring-communities-a-quick-start-guide-as \textgreater.

Bowman, S., Hyun, J., Perez, E., Chen, E., Pettit, C., Heiner, S., ...
\& Kaplan, J. (2022). Measuring progress on scalable oversight for large
language models. ArXiv preprint.
\textless https://arxiv.org/abs/2211.03540\textgreater.

Bradshaw, S., and Howard, P. N. (2018). Challenging truth and trust: A
global inventory of organized social media manipulation. \emph{Oxford
Internet Institute Computational Propaganda Project Report}.
\textless https://demtech.oii.ox.ac.uk/wp-content/uploads/sites/12/2018/07/ct2018.pdf\textgreater{}

Brundage, M., Avin, S., Wang, J., Belfield, H., Krueger, G., Hadfield,
G., Khlaaf, H., Yang, J., Toner, H., Fong, R., Maharaj, T., Koh, P. W.,
Hooker, S., Leung, J., Trask, A. Bluemke, E., Lebensold, J., O'Keefe,
C., Koren, M. \ldots{} and Anderljung, M. (2020). Toward trustworthy AI
development: mechanisms for supporting verifiable claims. ArXiv
preprint. \textless https://arxiv.org/abs/2004.07213\textgreater{}

Buolamwini, J. and Gebru, T. (2018). Gender shades: Intersectional
accuracy disparities in commercial gender classification.
\emph{Proceedings of the 1st Conference on Fairness, Accountability and
Transparency, (PMLR)} 81:77-91.

Cappelen, H. (2018). \emph{Fixing Language: An Essay on Conceptual
Engineering}. Oxford University Press.

Carlini, N., Athalye, A., Papernot, N., Brendel, W., Rauber, J.,
Tsipras, D., Goodfellow, I., Madry, A., and Kurakin, A. (2019). On
evaluating adversarial robustness. ArXiv preprint.
\textless https://arxiv.org/abs/1902.06705\textgreater{}

Carlsmith, J. (2021). Is power-seeking AI an existential risk? ArXiv
preprint. \textless https://arxiv.org/abs/2206.13353\textgreater{}

Chandola, V., Banerjee, A., and Kumar, V. (2009). Anomaly detection: A
survey. \emph{ACM computing surveys (CSUR)} 41: 1-58.

Chen, M. 2023. (2023). Artists and Illustrators Are Suing Three A.I. Art
Generators for Scraping and ``Collaging'' Their Work Without Consent'.
\emph{Artnet News} February 16, 2023.
\textless https://news.artnet.com/art-world/class-action-lawsuit-lensa-ai-prisma-labs-biometric-information-2257096\textgreater{}

Christiano, P. F., Leike, J., Brown, T., Martic, M., Legg, S., \&
Amodei, D. (2017). Deep reinforcement learning from human preferences.
\emph{Advances in neural information processing systems}, 30.

Clark, A. and Chalmers, D. (1998). The extended mind. \emph{Analysis}
58: 7--19.

Clark, J. and Amodei, D. (2016). Faulty reward function in the wild.
OpenAI blog post.
\textless https://openai.com/research/faulty-reward-functions\textgreater{}

Corso, A., Karamadian, D., Valentin, R., Cooper, M., and Kochenderfer,
M. J. (2023). A Holistic Assessment of the Reliability of Machine
Learning Systems. ArXiv preprint.
\textless https://arxiv.org/abs/2307.10586\textgreater{}

Costanza-Chock, S., Raji, I. D., and Buolamwini, J. (2022). Who Audits
the Auditors? Recommendations from a Field Scan of the Algorithmic
Auditing Ecosystem. \emph{Proceedings of the 2022 ACM Conference on
Fairness, Accountability, and Transparency (FAccT '22),} 1571--83.

Cotra, A. (2021). The Case for Aligning Narrowly Superhuman Models.
Alignment Forum Blog Post.
\textless https://www.alignmentforum.org/posts/PZtsoaoSLpKjjbMqM/the-case-for-aligning-narrowly-superhuman-models.\textgreater

Crawford, K. (2021). \emph{Atlas of AI: Power, Politics, and the
Planetary Costs of Artificial Intelligence.} Yale University Press.

Dafoe, A. (2018). AI Governance: A Research Agenda. \emph{Oxford
University Future of Humanity Institute Report}.
\textless https://www.fhi.ox.ac.uk/wp-content/uploads/GovAI-Agenda.pdf \textgreater.

D'Alessandro, W. (2024). Deontology and safe artificial intelligence.
\emph{Philosophical Studies}. Online First.

Dalrymple, David, Joar Skalse, Yoshua Bengio, Stuart Russell, Max
Tegmark, Sanjit Seshia, Steve Omohundro, et al. (2024) `Towards
Guaranteed Safe AI: A Framework for Ensuring Robust and Reliable AI
Systems'. arXiv, 8 July 2024. https://doi.org/10.48550/arXiv.2405.06624.

Dembroff, R. (2016). What is sexual orientation? \emph{Philosophers'
Imprint} 16: 1--27.

Dobbe, R. I. J. (2022). System Safety and Artificial Intelligence. In
Justin B. Bullock, Yu-Che Chen, Johannes Himmelreich, Valerie M. Hudson,
Anton Korinek, Matthew M. Young, and Baobao Zhang (eds). \emph{The
Oxford Handbook of AI Governance}, Oxford University Press.
\textless https://doi.org/10.1093/oxfordhb/9780197579329.013.67\textgreater{}

Elhage, N., Nanda, N., Olsson, C., Henighan, T., Joseph, N., Mann, B.,
... \& Olah, C. (2021). A Mathematical Framework for Transformer
Circuits. Blog Post.
\textless https://transformer-circuits.pub/2021/framework/index.html\textgreater{}

Fang, X. and Johnson, N. (2019). Three Reasons Why: Framing the
Challenges of Assuring AI. In Alexander Romanovsky, Elena Troubitsyna,
Ilir Gashi, Erwin Schoitsch, and Friedemann Bitsch (eds.) \emph{Computer
Safety, Reliability, and Security}, Springer, 281--87.

Feldstein, S. (2019). The Global Expansion of AI Surveillance.
\emph{Carnegie Endowment for International Peace Working Paper.}
\textless https://carnegieendowment.org/files/WP-Feldstein-AISurveillance\_final1.pdf\textgreater{}

Gabriel, I. (2020). Artificial intelligence, values, and alignment.
\emph{Minds and Machines}, 30(3), 411-437.

Ganguli, D., Lovitt, L., Kernion, J., Askell, A., Bai, Y., Kadavath, S.,
... \& Clark, J. (2022). Red teaming language models to reduce harms:
Methods, scaling behaviors, and lessons learned. ArXiv preprint.
\textless https://arxiv.org/abs/2209.07858\textgreater.

Geiger, A., Potts, C., and Icard, T. (2023). Causal Abstraction for
Faithful Model Interpretation. ArXiv Preprint.
\textless https://arxiv.org/abs/2301.04709\textgreater{}

Goldstein, S. and Kirk-Giannini, C. D. (2023). Language Agents Reduce
the Risk of Existential Catastrophe. \emph{AI \& Society}. Online First.

Grant, D. G., Behrends, J., and Basl, J. (2023). What We Owe to
Decision-Subjects: Beyond Transparency and Explanation in Automated
Decision-Making. \emph{Philosophical Studies}. Online First.

Green, Ben, and Lily Hu. (2018). The Myth in the Methodology:~Towards
Recontextualization of Fairness in Machine Learning. \emph{Presented at
the 35th International Conference on Machine Learning (ICML '18).}
\textless https://econcs.seas.harvard.edu/files/econcs/files/green\_icml18.pdf\textgreater.

Harding, J. (2023). Operationalising Representation in Natural Language
Processing. \emph{British Journal for the Philosophy of Science}.
\textless https://arxiv.org/abs/2306.08193\textgreater{}.

Harding, J. and Kirk-Giannini, C. D. (2023). AI's future worries us. So
does AI's present. \emph{Boston Globe} July 14, 2023.

Harding, J. and Sharadin, N. (Forthcoming). What is it for a Machine
Learning Model to Have a Capability? \emph{British Journal for the
Philosophy of Science.}

Haslanger, S. (1995). Ontology and Social Construction.
\emph{Philosophical Topics} 23: 95--125. Reprinted in Haslanger (2012),
pp. 83--112.

Haslanger, S. (2000). Gender and race: (What) are they? (What) do we
want them to be? \emph{Noûs} 34: 31--55. Reprinted in Haslanger (2012),
pp. 221--247.

Haslanger, S. (2005). What Are We Talking About? The Semantics and
Politics of Social Kinds. \emph{Hypatia} 20(4):10--26. Reprinted in
Haslanger (2012), pp. 365--380.

Haslanger, S. (2006). What Good Are Our Intuitions? Philosophical
Analysis and Social Kinds. \emph{Proceedings of the Aristotelian Society
Supplementary Volume} 80: 89--118. Reprinted in Haslanger (2012), pp.
381--405.

Haslanger, S. (2012). \emph{Resisting Reality: Social Construction and
Social Critique}. Oxford University Press.

Helbing, D., Frey, B.S., Gigerenzer, G., Hafen, E., Hagner, M.,
Hofstetter, Y., Van Den Hoven, J., Zicari, R.V., \& Zwitter, A. (2019).
Will democracy survive big data and artificial intelligence? In Helbing,
D. (ed.) \emph{Towards Digital Enlightenment: Essays on the Dark and
Light Sides of the Digital Revolution}, Springer, pp.73-98.

Hendrycks, D., Carlini, N., Schulman, J., and Steinhardt, J. (2022).
Unsolved Problems in ML Safety. ArXiv Preprint. \textless{}
https://arxiv.org/abs/2109.13916\textgreater{}

Hendrycks, D. and Dietterich, T. (2019). Benchmarking Neural Network
Robustness to Common Corruptions and Perturbations. \emph{International
Conference on Learning Representations 2019}.

Hendrycks, D., Mazeika, M., and Woodside, T. (2023). An Overview of
Catastrophic AI Risks. ArXiv preprint.
\textless https://arxiv.org/abs/2306.12001\textgreater{}

Hendrycks, D., Mazeika, M., Zou, A., Patel, S., Zhu, C., Navarro, J.,
Song, D., Li, B. and Steinhardt, J. (2021). What would Jiminy Cricket
do? Toward agents that behave morally. \emph{35th Conference on Neural
Information Processing Systems (NeurIPS 2021)}.

Howard, P. N., Woolley, S., \& Calo, R. (2018). Algorithms, bots, and
political communication in the US 2016 election: The challenge of
automated political communication for election law and administration.
\emph{Journal of Information Technology \& Politics} 15: 81--93.

E. Hutchins. How a Cockpit Remembers Its Speeds. Cognitive Science,
19(3):265-288, July 1995. ISSN 03640213. doi:
10.1207/s15516709cog1903\_1.

Irving, G., Christiano, P., \& Amodei, D. (2018). AI safety via debate.
ArXiv preprint. \textless https://arxiv.org/abs/1805.00899\textgreater.

Isaac, M. G. (2020). How to conceptually engineer conceptual
engineering? \emph{Inquiry}. Online First.

Kasirzadeh, A. (2024). Two Types of AI Existential Risk: Decisive and
Accumulative. ArXiv preprint.
\textless https://arxiv.org/abs/2401.07836\textgreater.

Keller, T. R., \& Klinger, U. (2019). Social bots in election campaigns:
Theoretical, empirical, and methodological implications. \emph{Political
Communication} 36: 171--189.

Khlaaf, H., Mishkin, P., Achiam, J., Krueger, G, and Brundage, M.
(2022). A Hazard Analysis Framework for Code Synthesis Large Language
Models. ArXiv preprint.
\textless https://arxiv.org/abs/2207.14157\textgreater{}

Kinniment, M., Sato L. J. K., Du, H., Goodrich, B., Hasin, M., Chan, L.,
Miles, L. H., Lin, T. R., Wijk, H., Burget, J., Ho, A., Barnes, E., and
Christiano, P. (2023). Evaluating Language-Model Agents on Realistic
Autonomous Tasks. \emph{Alignment Research Center Report}.
\textless https://evals.alignment.org/Evaluating\_LMAs\_Realistic\_Tasks.pdf\textgreater{}

Kirk-Giannini, C. D. (2023). Dilemmatic gaslighting. \emph{Philosophical
Studies}. Online First.

Kirk-Giannini, C. D. (Forthcoming). How to solve the gender inclusion
problem. \emph{Hypatia}.
\textless https://philpapers.org/archive/KIRHTS.pdf\textgreater.

Koessler, L., and Schuett, J. (2023). Risk Assessment at AGI Companies:
A Review of Popular Risk Assessment Techniques from Other
Safety-Critical Industries. ArXiv Preprint.
\textless https://arxiv.org/abs/2307.08823\textgreater{}

Langosco L., Koch, J., Sharkey, L., Pfau, J., and Krueger, D. (2022).
Goal misgeneralization in deep reinforcement learning. \emph{Proceedings
of the 39th International Conference on Machine Learning (ICML `22)}:
12004--12019.

Lazar, S. (2022). Power and AI: Nature and Justification. In \emph{The
Oxford Handbook of AI Governance}, edited by Justin B. Bullock, Yu-Che
Chen, Johannes Himmelreich, Valerie M. Hudson, Anton Korinek, Matthew M.
Young, and Baobao Zhang. Oxford University Press.

Lazar, Seth, and Alondra Nelson. (2023) `AI Safety on Whose Terms?'
\emph{Science} 381, no. 6654: 138--138.

Li, K., Patel, O., Viégas, F., Pfister, H., \& Wattenberg, M. (2024).
Inference-time intervention: Eliciting truthful answers from a language
model. \emph{Advances in Neural Information Processing Systems} 36.

Manheim, K., \& Kaplan, L. (2019). Artificial intelligence: Risks to
privacy and democracy. \emph{Yale Journal of Law and Technology} 21:
106--188.

Nemitz P. (2018). Constitutional democracy and technology in the age of
artificial intelligence. \emph{Philosophical Transactions of the Royal
Society A} 376: 20180089.

Ngo, R. (2022). AGI safety from first principles: Introduction. AI
Alignment Forum post.
\textless https://www.alignmentforum.org/s/mzgtmmTKKn5MuCzFJ/p/8xRSjC76HasLnMGSf\textgreater.

Ngo, R., Chan, L. and Mindermann, S. (2023). The Alignment Problem from
a Deep Learning Perspective (v5). ArXiv Preprint.
\textless https://arxiv.org/abs/2209.00626\textgreater{}

Noble, S. U. (2018). \emph{Algorithms of Oppression: How Search Engines
Reinforce Racism}. NYU Press.

Ord, T. (2020). \emph{The Precipice: Existential Risk and the Future of
Humanity.} Hachette Books.

Pang, G., Shen, C., Cao, L., \& Hengel, A. V. D. (2021). Deep learning
for anomaly detection: A review. \emph{ACM Computing Surveys (CSUR)}
54(2): 1-38.

Park, P. S., Goldstein, S., O\textquotesingle Gara, A., Chen, M., \&
Hendrycks, D. (2024). AI deception: A survey of examples, risks, and
potential solutions. \emph{Patterns} 5(5): 100988.

Perrigo, B. 2023. Exclusive: OpenAI Lobbied E.U. to Water Down AI
Regulation'. \emph{TIME Magazine} June 20, 2023.
\textless https://time.com/6288245/openai-eu-lobbying-ai-act/ \textgreater.

Rafailov, R., Sharma, A., Mitchell, E., Manning, C. D., Ermon, S., \&
Finn, C. (2024). Direct preference optimization: Your language model is
secretly a reward model. \emph{Advances in Neural Information Processing
Systems} 36.

Raji, I. D., Smart, A., White, R. M., Mitchell, M., Gebru, T.,
Hutchinson, B., Smith-Loud, J., Theron, D., and Barnes, P. (2020).
Closing the AI Accountability Gap: Defining an End-to-End Framework for
Internal Algorithmic Auditing. \emph{Proceedings of the 2020 Conference
on Fairness, Accountability, and Transparency}: 33--44.

Rismani, S., Shelby, R., Smart, A., Delos Santos, R., Moon, A. J., and
Rostamzadeh, N. (2023). Beyond the ML Model: Applying Safety Engineering
Frameworks to Text-to-Image Development. \emph{Proceedings of the 2023
AAAI/ACM Conference on AI, Ethics, and Society}: 70--83.

Rismani, S., Shelby, R., Smart, A., Jatho, E., Kroll, J., Moon, A. J.,
and Rostamzadeh, N. (2023). From Plane Crashes to Algorithmic Harm:
Applicability of Safety Engineering Frameworks for Responsible ML.
\emph{Proceedings of the 2023 CHI Conference on Human Factors in
Computing Systems}: 1--18.

Roland, H. and Moriarty, B. (1990). \emph{System Safety Engineering and
Management.} John Wiley \& Sons, Ltd.

Roose, K. (2023). A.I. Poses `Risk of Extinction,' Industry Leaders
Warn. \emph{New York Times} May 30, 2023.

Russell, S. (2019). \emph{Human Compatible: AI and the Problem of
Control}. Allen Lane.

Schneier, B. and Sanders, N. (2023). The A.I. Wars Have Three Factions,
and They All Crave Power. \emph{New York Times} September 28, 2023.

Selbst, A. D., Boyd, D., Friedler, S. A., Venkatasubramanian, S., and
Vertesi, J. (2019). Fairness and Abstraction in Sociotechnical Systems.
\emph{Proceedings of the Conference on Fairness, Accountability, and
Transparency (FAT* '19)}: 59--68.

Shavit, Y. (2023). What Does It Take to Catch a Chinchilla? Verifying
Rules on Large-Scale Neural Network Training via Compute Monitoring.
ArXiv preprint. \textless https://arxiv.org/abs/2303.11341\textgreater{}

Shelby, R., Rismani, S., Henne, K., Moon, A. J., Rostamzadeh, N.,
Nicholas, P., Yilla-Akbari, N. \ldots{} and Virk, G. (2023).
Sociotechnical Harms of Algorithmic Systems: Scoping a Taxonomy for Harm
Reduction. \emph{Proceedings of the 2023 AAAI/ACM Conference on AI,
Ethics, and Society}: 723--741.

Shevlane, T., Farquhar, S., Garfinkel, B., Phuong, M., Whittlestone, J.,
Leung, J., Kokotajlo, D., Marchal, N., Anderljung, M., Kolt, N. and Ho,
L. (2023). Model evaluation for extreme risks. ArXiv preprint.
\textless https://arxiv.org/abs/2305.15324\textgreater{}

Solaiman, I., Talat, Z., Agnew, W., Ahmad, L., Baker, D., Blodgett, S.
L., ... \& Vassilev, A. (2023). Evaluating the Social Impact of
Generative AI Systems in Systems and Society. ArXiv preprint.
\textless https://arxiv.org/abs/2306.05949\textgreater{}

Tanswell, F. S. (2018). Conceptual engineering for mathematical
concepts. \emph{Inquiry} 61: 881-913.

Thornley, E. (2024). The shutdown problem: an AI engineering puzzle for
decision theorists. \emph{Philosophical Studies}. Online First.

Trapp, M., Schneider, D., and Weiss, G. (2018). Towards Safety-Awareness
and Dynamic Safety Management. \emph{14th European Dependable Computing
Conference (EDCC)}, 107--11.

Trowler, P. (2012). Disciplines and academic practices. In Trowler, P.,
Saunders, M., and Bamber, V. (2012), pp. 30--38.

Trowler, P., Saunders, M., and Bamber, V. (eds) (2012). \emph{Tribes and
Territories in the 21st Century: Rethinking the Significance of
Disciplines in Higher Education.} Routledge.

Tubert, A., \& Tiehen, J. (2024). Existentialist risk and value
misalignment. \emph{Philosophical Studies}. Online First.

Verma, S., Boonsanong, V., Hoang, M., Hines, K. E., Dickerson, J. P.,
and Shah, C. (2022). Counterfactual Explanations and Algorithmic
Recourses for Machine Learning: A Review. ArXiv Preprint.
\textless https://arxiv.org/abs/2010.10596\textgreater{}

Weidinger, L., Mellor, J., Rauh, M., Griffin, C., Uesato, J., Huang, P.
S., ... \& Gabriel, I. (2021). Ethical and social risks of harm from
language models. ArXiv preprint.
\textless https://arxiv.org/abs/2112.04359\textgreater{}

Weidinger, L., Rauh, M., Marchal, N., Manzini, A., Hendricks, L. A.,
Mateos-Garcia, Bergman, S. \ldots{} and Isaac, W. (2023). Sociotechnical
Safety Evaluation of Generative AI Systems. ArXiv Preprint.
\textless https://arxiv.org/abs/2310.11986\textgreater{}

Weidinger, L., Uesato, J., Rauh, M., Griffin, C., Huang, P. S., Mellor,
J., ... \& Gabriel, I. (2022). Taxonomy of risks posed by language
models. \emph{Proceedings of the 2022 ACM Conference on Fairness,
Accountability, and Transparency}: 214-229.

Wijk, H., Lin, T., Becker, J., Jawhar, S., Parikh, N., Broadley, T., ...
\& Barnes, E. (2024). RE-Bench: Evaluating frontier AI R\&D capabilities
of language model agents against human experts. ArXiv preprint.
\textless https://arxiv.org/abs/2411.15114\textgreater.

Yampolskiy, R., and Fox, J. (2013). Safety Engineering for Artificial
General Intelligence. \emph{Topoi} 32: 217--26.

Yang, J., Zhou, K., Li, Y., \& Liu, Z. (2021). Generalized
out-of-distribution detection: A survey. ArXiv Preprint.
\textless https://arxiv.org/abs/2110.11334\textgreater.

}


\end{document}